# Bound States in the Continuum Based on the Total Internal Reflection of Bloch Waves

Peng Hu[1,†], Chongwu Xie[1,†], Qianju Song[1], Ang Chen[2], Hong Xiang[1,3], Dezhuan Han[1,*], and Jian Zi[2,*]

[1] College of Physics, Chongqing University, Chongqing 401331, China

[2] State Key Laboratory of Surface Physics, Key Laboratory of Micro- and Nano-Photonic Structures (Ministry of Education) and Department of Physics, Fudan University, Shanghai 200433, China

[3] Chongqing Key Laboratory for Strongly Coupled Physics, Chongqing 401331, China

[*]**Corresponding authors.** Email: dzhan@cqu.edu.cn and jzi@fudan.edu.cn
[†]Equally contributed to this work.

## ABSTRACT

A photonic-crystal slab can support bound states in the continuum (BICs) which have infinite lifetimes but embedded into the continuous spectrum of optical modes in free space. The formation of BICs requires a total internal reflection (TIR) condition at both interfaces between the slab and free space. Here, we show that the TIR of Bloch waves can be directly obtained based on the generalized Fresnel equations proposed. If each of these Bloch waves picks up a phase with integer multiples of $2\pi$ for traveling a round trip, light can be perfectly guided in the slab, namely, forming a BIC. A BIC solver with low computational complexity and fast convergence speed is developed, which can also work efficiently at high frequencies beyond the diffraction limit where multiple radiation channels exist. Two examples of multi-channel BICs are shown, and their topological nature in momentum space is also revealed. Both can be attributed to the coincidence of the topological charges of far-field radiations from different radiation channels. The concept of the generalized TIR and the TIR-based BIC solver developed offer highly effective approaches for explorations of BICs which could have many potential applications in guided-wave optics and enhanced light-matter interactions.

**Keywords:** bound states in the continuum, total internal reflection, Bloch waves, generalized Fresnel equations, topological charges

# INTRODUCTION

Bound states in the continuum (BICs) are a special kind of resonant states with infinite lifetimes even though they are embedded into the continuous spectrum of free space [1-3], originally proposed by von Neumann and Wigner for an electron in a specially designed local potential [4]. Recently, BICs have been found to be a generic wave phenomenon existing in various physical systems, such as photonic [5-22], acoustic [23], and plasmonic ones [24, 25]. The ultra-high Q factors near BICs (including quasi-BICs [26, 27]) render many interesting applications possible, such as polarization control [11, 12], lasing [28-31], sensing [32], and non-linear optics [27].

As a platform for nanophotonics, photonic-crystal (PhC) slabs can guide light perfectly for optical modes below the light cone [33]. Above the light cone, guided modes become guided resonances since they are leaky [34, 35]. BICs can exist as isolated points on the bands of guide resonances [5-13]. From the far-field viewpoint, they can be interpreted as the vortex singularities of far-field polarizations with quantized topological charges [9]. These topological charges can be created, annihilated, and merged in the Brillouin zone [12-15]. From the viewpoint of wave interference, some BICs in PhC slabs can be treated as the Friedrich–Wintgen type which originates from the destructive interference of two different guided resonances [36-38].

We proposed that the formation mechanism of BICs in a PhC slab can be further interpreted in terms of the interference of bulk Bloch states [10]. For a uniform dielectric slab, the formation of guided waves require two conditions: a total internal reflection (TIR) at the interfaces between the slab and free space and that waves along the direction perpendicular to the slab are standing ones. The formations of BICs in a PhC slab must also satisfy these two conditions. In a uniform dielectric slab, the condition of TIR is simply that the angle between the propagating direction and the slab-surface normal is greater than the critical angle. However, any optical mode supported in a PhC slab is the superposition of bulk Bloch waves of such an infinite PhC rather than single plane wave. As a result, the TIR condition for a PhC slab becomes that of the TIR of constituent Bloch waves. If the total transmission for an optical mode consisting of multiple Bloch waves from the PhC slab side to the free-space side vanish, TIR will occur. It is just the condition of the TIR of Bloch waves stemming from the multiple interference of the constituent Bloch waves. Therefore, the study of BICs can start from a basic problem: the diffraction of Bloch waves at a single interface. The key point is that there may exist multiple reflected and refracted waves because of Bragg scattering [33, 39]. At the interface between a uniform dielectric and free space, the wave vector component $\mathbf{k}_\parallel$ parallel to the interface is a good quantum number due to the continuous translational symmetry at the interface. TIR can occur when $|\mathbf{k}_\parallel|$ is larger than the free-space wave vector since the perpendicular component of the wave vector $k_\perp$ on the free-space side becomes imaginary. In PhCs, the continuous translational symmetry is broken. However, the discrete translational symmetry leads to the equivalence of $\mathbf{k}_\parallel$ and $\mathbf{k}_\parallel + n\mathbf{G}$, where $n$ is an integer and $\mathbf{G}$ is a reciprocal lattice vector. This new degree of freedom renders the TIR of Bloch waves possible via a coherent way [10, 21] to be discussed in detail later.

Here, the TIR of Bloch waves is fully investigated from the viewpoint of diffraction. The generalized Fresnel equations for Bloch waves are derived, and formulas for the TIR of two Bloch waves with a very compact form are obtained analytically. For PhC slabs, the conventional conditions for the existence of waveguide modes can be directly generalized based on the TIR of Bloch waves, and the solutions of the generalized conditions are exactly BICs. A BIC solver is therefore developed with low computational complexity and fast convergence speed, and can be used for the search and determination of BICs in a very large parameter space. Different from previous studies of BICs in PhC slabs, which are restricted to a single radiation channel, the generalized conditions can be also applied to the case of multiple radiation channels. Therefore, the BIC solver can find BICs for any number of radiation channels at any high frequency. Examples of BICs with two radiation channels are given, and it is demonstrated that multi-channel BICs require the coincidence of the topological charges of far-filed radiations in all radiation channels.

## THEORY AND RESULTS

### Theory for the total internal reflection of Bloch waves

The TIR of Bloch waves can be interpreted from the perspective of diffraction. We start from a simple planar grating shown in Fig. 1(a). If there is only 0th diffraction order, the direct transmission is not zero in general. Therefore the simplest non-trivial case is that there are two propagating diffraction orders with wave vectors $\mathbf{k}$ and $\mathbf{k} - \mathbf{G}$. It is the discrete translational symmetry of the grating that leads to the Bragg scattering between $\mathbf{k}$ and $\mathbf{k} + n\mathbf{G}$, offering more degree of freedom to control the incident waves which is not restricted to a single plane wave with a fixed $\mathbf{k}$. To be specific, if a zero total transmission can occur by introducing two incident plane waves with **k** and **k – G**, TIR is thus realized in a very unusual way even though the transmission for each of the incident wave is not zero. In Fig. 1(a), diffraction of incident plane waves with wave vector **k** (purple arrows) and **k – G** (red arrows) are shown. For both cases, there are 0th and –1st (or 1st) diffraction orders. By definition, the 0th order remains the same wave vector as the incident one, so the order of the principal maximum and the secondary maximum is in fact switched for these two incident waves. If the two plane waves, **k** and **k – G**, are incident onto the grating simultaneously, the elimination of transmission shown in the right figure of Fig. 1(a) requires that the intensities of the diffracted waves should satisfy the relation:

$$I_{\mathbf{k}}(-1)/I_{\mathbf{k}}(0) = I_{\mathbf{k-G}}(0)/I_{\mathbf{k-G}}(1), \tag{1}$$

where $I_{\mathbf{k}}(m)$ represents the intensity of the $m$th-order diffraction for the incident wave vector **k**. However, according to the Fraunhofer diffraction from a diffraction grating [40], the principal maximum of $m = 0$ is usually the dominant maximum, and Eq. (1) cannot be satisfied generally.

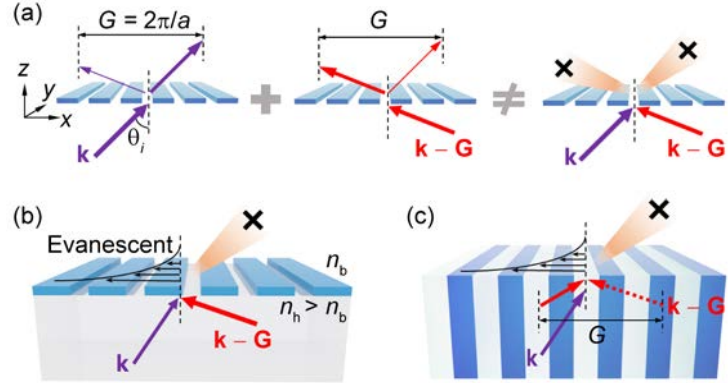

**FIG. 1.** (a) Diffraction of a simple planar grating with only two diffraction orders. For the incident plane wave with a wave vector **k** (or **k – G**), the intensity of the transmitted wave has a principal maximum in the direction of **k** (or **k – G**), and a secondary maximum in the direction of **k – G** (or **k**). Therefore, total reflection cannot be realized by altering the relative coefficient of incident waves **k** and **k – G** to form destructive interference in both diffraction orders on the transmission side. (b) Combination of a high index ($n_h$) medium and a planar grating. The introduction of a high index medium can convert the propagating diffraction order with $\mathbf{k}_b - \mathbf{G}$ to be evanescent, where $\mathbf{k}_b$ is the wave vector in the background medium with a refractive index $n_b$, and $\mathbf{k}_{b,\parallel} = \mathbf{k}_\parallel$. Only one propagating diffraction order survives on the transmission side under the condition $n_b k_0 < |\mathbf{k}_\parallel - \mathbf{G}| < n_h k_0$. (c) A semi-infinite PhC acting as the combination of a high index medium and a grating. Incident waves now should be changed from plane waves to Bloch waves.

If a high index ($n_h$) material is adopted as a substrate shown in Fig. 1(b), we can possibly make the propagating diffraction order with wave vector $\mathbf{k}_b - \mathbf{G}$ evanescent. Here, $\mathbf{k}_b$ is the wave vector on the transmission side with refractive index $n_b$, and $\mathbf{k}_{b,\parallel} = \mathbf{k}_\parallel$. The condition is that $\mathbf{k}_\parallel$ satisfies $n_b k_0 < |\mathbf{k}_\parallel - \mathbf{G}| < n_h k_0$, where $k_0$ is the free-space wave vector. Under this condition, only one propagating diffraction order survives for both the incidence of $\mathbf{k}$ and $\mathbf{k} - \mathbf{G}$, and Eq. (1) is relaxed greatly and reduced to $I_\mathbf{k}(0) = I_{\mathbf{k}-\mathbf{G}}(1)$. The destructive interference of the transmitted waves can be readily realized just by appropriately choosing the relative phase and amplitude of the two incident waves. Note that the essential point is that we have a sufficient degree of freedom for the incident waves to cancel out the transmission. Similar destructive interference was considered to achieve some unique phenomena such as complete reflections [41-43] and perfect anti-reflections [44].

In fact, the combination of a grating and a high index material can be replaced by a PhC [33], as shown in Fig. 1(c). We focus on a one-dimensional semi-infinite PhC with a period of $a$ in the $x$ direction and uniform in the $y$ direction. The alternating dielectric layers in the PhC have relative permittivity $\varepsilon_1$ and $\varepsilon_2$, and thicknesses $a - d$ and $d$, respectively. The background is chosen to be air with $\varepsilon_b = 1$. Different from that in the Fraunhofer diffraction of gratings, we choose the incident waves as Bloch waves rather than plane waves since Bloch waves are eigenstates of the periodic structure and any optical modes supported are a superposition of these Bloch waves. A Bloch wave with wave vector $\mathbf{k} + n\mathbf{G}$ is equivalent to that with $\mathbf{k}$. Obviously, all of the arguments above for the existence of TIR can be applied to Bloch waves here.

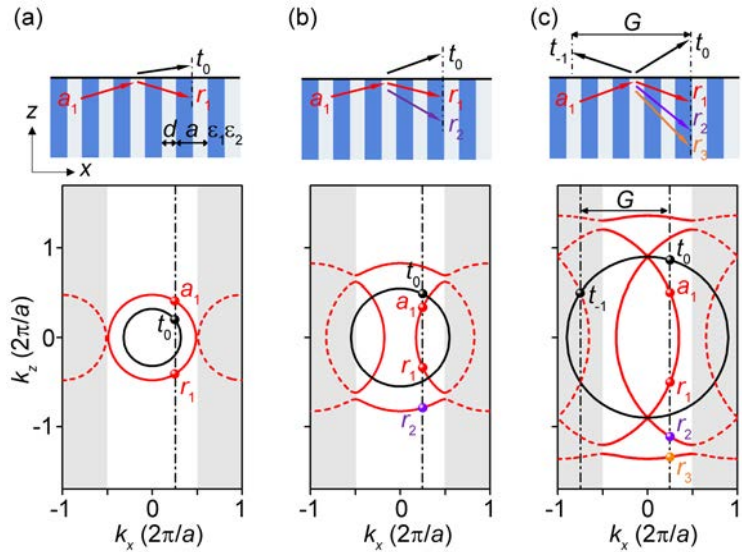

**FIG. 2.** (a)–(c) Diffraction of a single Bloch wave incident from a PhC to air shown in the upper panels and corresponding isofrequency contours in the lower panels for three examples from low to high frequency. The isofrequency contours for the PhC in (out of) the first Brillouin zone are denoted by solid (dashed) red lines, whereas the isofrequency contours in air are shown by black lines. The parallel wave vectors of the incident, reflected, and transmitted waves are the same and indicated by black dashed lines. The number of propagating Bloch waves in the PhC ($N_p$) and that of propagating diffraction orders in air ($N_r$) for a fixed $k_x$ are $N_p = 1$ and $N_r = 1$ in (a), $N_p = 2$ and $N_r = 1$ in (b), and $N_p = 3$ and $N_r = 2$ in (c).

The existence of multiple Bloch waves can be clearly seen in isofrequency contours. The dispersion relation, which relates the frequency $\omega$, the normal wave vector $k_z$, and the Bloch wave vector $k_x$ is given in the Supplementary Information. Fig.2 shows three examples with different frequencies $\omega$. The isofrequency contours in air are shown by black lines, whereas those for the PhC are indicated by red lines. The red lines will be folded back when they go beyond the first Brillouin zone (see the dashed red lines) due to the periodicity in the $x$ direction. In Fig. 2(a), at a low frequency, the isofrequency contours for the PhC and air in the first Brillouin zone are simply two circles without band folding, offering conventional refraction and transmission. The coefficient for the incident wave is denoted by $a_1$, whereas that for the reflected and transmitted waves are denoted by $r_1$ and $t_0$, respectively. The subscript in $t_0$ stands for 0th-order diffraction in air. In this case, the number of propagating Bloch waves in the PhC ($N_p$) and that of propagating diffraction orders in air ($N_r$) are equal to one. The propagating diffraction orders can also be viewed as radiation channels. As frequency increases, band folding takes place and band gaps appear at the edges of the first Brillouin zone, as shown in Fig. 2(b), and $N_p$ of the propagating Bloch waves is increased to two, whereas $N_r$ of the radiation channels is still one. Thus, for a single Bloch wave incident with a coefficient $a_1$, in addition to the reflection $r_1$ (the same Bloch wave), an additional Bloch wave with a coefficient $r_2$ will also be excited. When frequency is further increased, more Bloch waves will be present as additional reflected waves, such as $r_3$ shown in Fig. 2(c). Moreover, when frequency goes beyond the diffraction limit, the −1st-order diffracted wave in air will change from evanescent to propagating, so that for the case in Fig. 2(c), we have $N_p = 3$ and $N_r = 2$.

The formalism for the diffraction of Bloch waves incident from a PhC to air are outlined as follows. Here, we only consider transverse electric (TE) Bloch waves ( $\mathbf{E} = E_y \hat{y}, \mathbf{H} = H_x \hat{x} + H_z \hat{z}$ ). Transverse magnetic (TM) Bloch waves ( $\mathbf{H} = H_y \hat{y}, \mathbf{E} = E_x \hat{x} + E_z \hat{z}$ ) are discussed in the Supplementary Information. Suppose that a series of TE Bloch waves with a fixed frequency $\omega$ and Bloch wave vector $k_x$ impinge on the PhC/air interface at $z=0$. The electric field inside the PhC can be written as follows:

$$E_y^{\text{in}}(x,z) = \sum_{n=1}^{\infty} \left( a_n e^{ik_z^{(n)}z} + r_n e^{-ik_z^{(n)}z} \right) u^{(n)}(x) e^{ik_x x}, \qquad (2)$$

where $a_n$ and $r_n$ are respectively the complex coefficients of the incident and reflected Bloch waves, $k_z^{(n)}$ is the normal wave vector of the $n$th Bloch wave and $u^{(n)}(x)$ is the periodic-in-cell part of the $n$th Bloch wave function. Suppose we have $N_\text{p}$ incident propagating Bloch waves $1 \leq n \leq N_\text{p}$. Bloch waves with the order $n > N_\text{p}$ are evanescent waves with $k_z$ being purely imaginary. Physically, incident evanescent waves that increase away from the interface should be excluded for this semi-infinite PhC, namely, $a_{n>N_\text{p}} = 0$. The transmitted wave in air can be expressed as follows:

$$E_y^{\text{out}}(x,z) = \sum_{m=-\infty}^{\infty} t_m e^{i(k_{x,m}x + k_{z,m}z)}, \qquad (3)$$

where $k_{x,m} = k_x + mG$ and $k_{z,m} = \sqrt{k_0^2 - k_{x,m}^2}$. Here, $t_m$ is the complex transmission coefficient for the $m$th diffraction order. At the interface, a Fourier transform of the boundary conditions (the continuity of tangential $\mathbf{E}$ and $\mathbf{H}$ fields) gives:

$$\mathbf{T} = \vec{\mathbf{X}}(\mathbf{A} + \mathbf{R}) \qquad (4)$$

and

$$\vec{\Pi}\, \mathbf{T} = \vec{\Omega}(\mathbf{A} - \mathbf{R}), \qquad (5)$$

where $(\mathbf{T})_m = t_m$, $(\mathbf{A})_n = a_n$, $(\mathbf{R})_n = r_n$, $\vec{\mathbf{X}}_{mn} = \frac{1}{a}\int_{x_0}^{x_0+a} u^{(n)}(x) e^{-imGx} dx$, $\vec{\Omega}_{mn} = k_z^{(n)} \vec{\mathbf{X}}_{mn}$, and $\vec{\Pi}_{mn} = k_{z,m} \delta_{mn}$. Note that the first $N_\text{r}$ elements of $\mathbf{T}$ correspond to the radiation channels in air. To solve Eqs. (4) and (5), the number of Fourier components (indexed by $m$) should be chosen to be the same as the number of Bloch waves (indexed by $n$). By eliminating $\mathbf{T}$, the relation between the reflection and incidence reads:

$$\mathbf{R} = \left(\vec{\Omega} + \vec{\Pi}\vec{\mathbf{X}}\right)^{-1} \left(\vec{\Omega} - \vec{\Pi}\vec{\mathbf{X}}\right) \mathbf{A} \equiv \vec{\Phi} \mathbf{A}. \qquad (6)$$

Then, the transmission can be expressed as follows:

$$\mathbf{T} = \vec{\mathbf{X}}\left(\vec{\mathbf{I}} + \vec{\mathbf{\Phi}}\right)\mathbf{A} \equiv \vec{\mathbf{M}}\mathbf{A}, \tag{7}$$

where $\vec{\mathbf{I}}$ is the identity matrix. Since we are only interested in the transmission of the radiation channels (denoted by $\mathbf{T}_r$), Eq. (7) can be reduced to

$$\mathbf{T}_r = \vec{\mathbf{M}}_r \mathbf{A}_p, \tag{8}$$

where $\mathbf{A}_p$ only consists of the first $N_p$ items of $\mathbf{A}$, corresponding to the propagating Bloch waves, and $\vec{\mathbf{M}}_r$ is a submatrix of $\vec{\mathbf{M}}$ with elements $\vec{\mathbf{M}}_{ij}$ only taking $1 \leq i \leq N_r$ and $1 \leq j \leq N_p$.

Equations (6) and (8) are in fact the generalized Fresnel equations for Bloch waves. Based on these two equations, the problem of the incidence of any number of Bloch waves can be solved. Obviously, the TIR condition of Bloch waves is also a direct consequence, given by:

$$\mathbf{T}_r = 0. \tag{9}$$

When the number of propagating Bloch waves is equal to that of radiation channels, namely, $N_p = N_r$, a non-trivial solution of this condition requires that $\det\left(\vec{\mathbf{M}}_r\right) = 0$, which is difficult to realize for a PhC. However, if $N_p > N_r$, a non-trivial solution of incidence $\mathbf{A}_p$ always exists for Bloch waves.

Based on the TIR of Bloch waves, light can be further guided in a PhC slab with a finite thickness $h$. Distinct from the semi-infinite PhC, all evanescent Bloch waves are allowed in a PhC slab, with either positive or negative attenuation in the $z$ direction. The origin of the $z$ axis is now set at the center of the PhC slab for convenience. Equations (6) and (7) are also slightly modified via replacing $a_n$ with $a_n e^{-ik_z^{(n)}h}$. Supposing that the TIR condition, $\mathbf{T}_r = 0$, is satisfied at the upper interface for some properly initiated incidence $a_n$, the reflected waves will then become the incident waves at the lower interface. In the case that the ratio $r_n/a_n$ remains a constant for any arbitrary $n$, the TIR condition can be maintained at the lower interface.

However, the TIR condition is not the only condition for forming a waveguide mode. The phase accumulated after a round trip should be integer multiples of $2\pi$, also called the guidance condition [45]. This guidance condition can be directly generalized just by counting the accumulated phase for each Bloch wave. At the interface of the PhC slab, a phase shift $\varphi_r^{(n)} = \arg\left(r_n / a_n e^{-ik_z^{(n)}h}\right)$ takes place for the $n$th Bloch wave. Note that the additional term $e^{-ik_z^{(n)}h}$ in the phase shift comes from the shift of the origin of $z$ compared with the above semi-infinite PhC case. Similar to that for conventional waveguide modes, the total phase change for a round trip should be integer multiples of $2\pi$ for the $n$th Bloch wave, which can be simply expressed as follows:

$$k_z^{(n)}h + \varphi_r^{(n)} = m^{(n)}\pi, \tag{10}$$

where $m^{(n)}$ is an integer. Equations (9) and (10) can be viewed as the generalized conditions for waveguide modes in a PhC slab, as summarized in Table 1. Waveguide modes that satisfy the generalized conditions inside the light cone are precisely BICs. Combining Eqs. (9) and (10), we also obtain that $r_n / a_n = \pm 1$ for all Bloch waves, where the positive and negative signs correspond to even and odd symmetries in the $z$ direction, respectively.

**Table 1.** Conventional and generalized conditions for waveguide modes.

| Waveguide modes | TIR | Guidance condition |
| --- | --- | --- |
| Conventional | $\theta > \theta_c$ | $k_z h + \varphi_r = m\pi$ |
| Generalized | $t_i = 0$ for $1 \leq i \leq N_r$ | $k_z^{(n)}h + \varphi_r^{(n)} = m^{(n)}\pi$ |

The generalized conditions for waveguide modes can be used to efficiently determine BICs in the $k_x$–$\omega$ space. In addition to propagating Bloch waves, evanescent waves with purely imaginary $k_z$ can exist near the interface of the PhC slab and should also be taken into account. Based on the generalized conditions for waveguide modes, a BIC solver has been designed [46] with the advantage of very low computational complexity and fast convergence speed. Since the Bloch waves we adopt form an appropriate basis set inside the PhC, the positions of BICs in the $k_x$–$\omega$ space converge very quickly if only a few evanescent waves are considered, in addition to the propagating Bloch waves (see Supplementary Fig. 1). To be specific, we first ensure that the TIR condition is satisfied at one of the interfaces for every $(k_x, \omega)$ point. The TIR condition requires that the number of propagating Bloch waves is larger than that of radiation channels ($N_p > N_r$). The corresponding phase shift, $\varphi_r^{(n)}$, at this interface can be obtained by solving Eqs. (6) and (9) (see Supplementary Information for details). Second, we build a database of $\varphi_r^{(n)}$ for a PhC in the whole $k_x$–$\omega$ space. Finally, for any thickness $h$, the total phase of a round trip for the $n$th Bloch wave inside the PhC slab is simply $k_z^{(n)}h + \varphi_r^{(n)}$. What the solver should do is to determine whether this phase is integer multiples of $\pi$. Therefore, the computational time is mainly spent on the construction of a reflection-phase database. With this database, the time to search BICs for different $h$ values is negligible. We show an example of searching BICs in a range of $k_x$–$\omega$ space with $N_p = 2$ and $N_r = 1$ in Supplementary Fig. 1, in full agreement with the results simulated by the finite element method.

It is worth mentioning that the algorithm based on the generalized conditions for waveguide modes can be applied to not only the $k_x$ axis but the whole Brillouin zone. As a result, this BIC solver can work in the whole $\mathbf{k}_\parallel$–$\omega$ space, where $\mathbf{k}_\parallel = (k_x, k_y)$. However, BICs usually exist on high symmetry lines. In Supplementary Fig. 2, we give another example of searching BICs in the $k_y$ axis in the parameter space.

## TIR of two propagating Bloch waves

The simplest case for TIR of Bloch waves is that there are only two propagating Bloch waves in the PhC ($N_p = 2$) and one radiation channel in air ($N_r = 1$), as shown in Fig. 1(c). In this case, an analytical solution can be obtained. We assume that only propagating Bloch waves are considered and other evanescent waves are neglected. According to the above analysis, we only need to achieve the TIR of two Bloch waves at a single interface and then adopt the generalized guidance condition to fix BICs. The generalized Fresnel equations for Bloch waves can be simplified considerably and a concise form for the relative coefficient of the incident waves can be directly obtained when TIR occurs at the interface. For TE waves, it can be expressed as (see Supplementary Information for details):

$$\frac{a_2}{a_1} = -\frac{1+Z_2}{1+Z_1}, \tag{11}$$

where $Z_n = k_{z,-1}/k_z^{(n)}$ has a similar form of relative surface impedance [47]. Here, $k_{z,-1}$ is the normal wave vector of $-1$st-order diffracted wave in air, which is purely imaginary. Moreover, the reflection coefficients at the interface are as follows:

$$\frac{r_n}{a_n} = \frac{1-Z_n}{1+Z_n}, \tag{12}$$

which takes a similar form of the reflection coefficient in the conventional Fresnel equations. Note that Eq. (12) holds only when the TIR of two Bloch waves occurs.

The TIR condition becomes slightly complicated for two TM Bloch waves since the electric field is a vector in nature for the TM case but is a scalar for the TE case [48]. An approximate form of the Fourier transform of $\varepsilon(x)$ is used: $\varepsilon^{-1}(x) \sim \kappa_0 + \kappa_1 e^{iGx} + \kappa_{-1} e^{-iGx}$. When the TIR of two TM Bloch waves occurs, similar forms of the relative incidence and reflection coefficients can be obtained as those in Eqs. (11) and (12). However, the definition of $Z_n$ should be modified and expressed as follows (see Supplementary Information for details):

$$Z_1 = \frac{k_{z,-1}/\varepsilon_b}{k_z^{(1)}/\varepsilon_H} \text{ and } Z_2 = C\frac{k_{z,-1}/\varepsilon_b}{k_z^{(2)}/\varepsilon_H}, \tag{13}$$

where $\varepsilon_H = \kappa_0^{-1}$ and $C = \left(\left(k_z^{(2)}\right)^2 + k_x^2 - qG\right) \Big/ \left(\left(k_z^{(2)}\right)^2 + k_x^2 + k_x G - G^2\right)$.

The TIR condition of two Bloch waves can be realized via Eqs. (11) and (12) with appropriate definitions of $Z_n$ for TE and TM waves. The phase shift of TIR for the $n$th Bloch wave is as follows:

$$\varphi_r^{(n)} = \arg(r_n/a_n) = 2\arctan(iZ_n). \tag{14}$$

Note that diffracted evanescent waves are neglected in the above TIR condition, so Eqs. (11-14) work for the case when the index contrast is not too large, namely, $\Delta = |\varepsilon_2 - \varepsilon_1|/\varepsilon_1 \ll 1$. Strikingly, even when $\Delta \to 0$, i.e., the index contrast is vanishingly small, $Z_n$ approaches a constant for any $(k_x, \omega)$ point. Therefore, two important conclusions can be drawn: (1) BICs obtained from Eqs. (10) and (14) approach a series of fixed points in the $\mathbf{k}_\parallel$–$\omega$ space [21]. Generally, the band of guided resonances can be regarded as the folded band of the waveguide modes in an effectively uniform waveguide. The existence of discrete BIC points in the limiting case manifests the non-trivial physical consequence that continuous translational symmetry is broken into a discrete translational symmetry even if $\Delta \to 0$. (2) It is known that the introduction of a substrate can destroy BICs [49]. This is because the TIR conditions at the upper and lower interfaces are different, the combination of TIR at a single interface and guidance condition in Table 1 cannot restore the waveguide mode after a round trip. Or, in other words, the TIR conditions at the two interfaces contradict each other if there is a substrate.

**Multi-channel BICs**

When frequency increases, more than one propagating diffraction order (i.e. radiation channel) in air appears, shown in Fig. 2(c). The construction of BICs is more subtle since all radiation channels should be closed. Note that multi-channel BICs occurring at $k_x = 0$ or $\pi/a$ were discussed in Ref. [50]. However, since these multi-channel BICs appear at the high-symmetry points in the Brillouin zone, the corresponding radiation channels are not completely independent. Strikingly, the above generalized conditions for waveguide modes can be directly applied to the case with multiple radiation channels. The BIC solver we designed can thus work well to determine multi-channel BICs. Two different types of BICs with two radiation channels are taken as examples and shown in Figs. 3(a) and 3(b), which consist respectively of three and four propagating Bloch modes. These two multi-channel BICs appear on the $\text{TE}_0^{(1)}$ and $\text{TE}_0^{(-2)}$ bands, as highlighted by red dots in Fig. 3. Here, $\text{TE}_0^{(m)}$ represents the fundamental TE mode with $m$ being the index of band folding in the reduced-zone scheme. It is known that BICs interpreted by topological vortexes can exist robustly in the parameter space [9]. The robustness of BICs should be reexamined for multi-channel BICs since they only exist for some specific thicknesses, for example, $h_{\text{BIC}} = 1.948a$ and $1.968a$ in Figs. 3(a) and 3(b), respectively. The $Q$ factors of the guided resonance modes near the multi-channel BICs are plotted in Fig. 3(c) and (d) for different $h$ values. It can be clearly seen that the $Q$ factor diverges only when the thickness is equal to $h_{\text{BIC}}$. This divergence behavior disappears as long as the thickness is slightly varied away from $h_{\text{BIC}}$, which is distinct from robust BICs below the diffraction limit. The divergence rates are also plotted in Figs. 3(e) and 3(f), which are $Q \sim 1/\delta k_x^2$ and $Q \sim 1/\delta h^2$ (inverse square law), respectively. Here, $\delta k_x = |k_x - k_{x,\text{BIC}}|$ and $\delta h = |h - h_{\text{BIC}}|$.

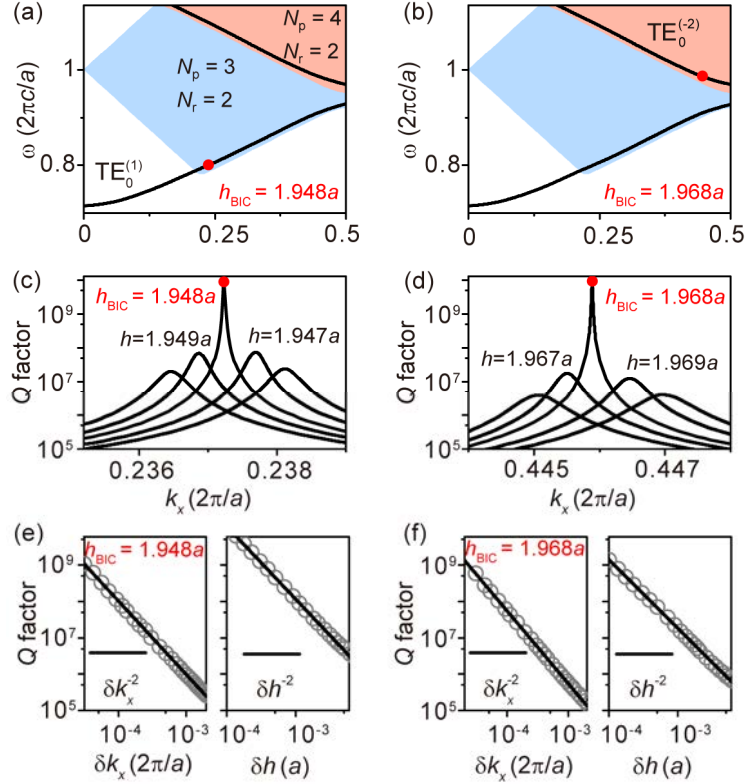

**FIG. 3.** Multi-channel BICs with two radiation channels. (a) and (b) Simulated band structures for $h=1.948a$ and $h=1.968a$, respectively. Multi-channel BICs (red dots) exist on the $TE_0^{(1)}$ band in (a) and $TE_0^{(-2)}$ band in (b). The light blue (orange) shaded region indicates the region in which there are three (four) propagating Bloch modes in the PhC and two radiation channels in air. (c) and (d) Simulated $Q$ factors of the guided resonances for different $h$ in the $TE_0^{(1)}$ and $TE_0^{(-2)}$ bands, respectively. (e) and (f) Divergence behavior of $Q$ factors for the two multi-channel BICs with $\delta k_x = |k_x - k_{x,\text{BIC}}|$ and $\delta h = |h - h_{\text{BIC}}|$. The solid lines in (e) and (f) represent the fitting of the inverse square of $\delta k_x$ or $\delta h$. Here, the other system parameters are chosen as $\varepsilon_1 = 1$, $\varepsilon_2 = 4.9$, and $d = 0.5a$.

It has been demonstrated that BICs below the diffraction limit are vortex centers of the polarization directions of far-field radiations [9], characterized by topological charges and robust then in parameter space. However, for multi-channel BICs, an increased number of radiation channels can make an essential difference and the topological nature is manifested in other ways. To reveal the topological nature, the far-field polarization states are investigated for each radiation channel (see Supplementary Information for details). The far-field polarization states displayed in Fig. 4(c) correspond to the multi-channel BIC shown in Fig. 3(a). There are two radiation channels in air and three propagating Bloch waves in the PhC, as shown in Fig. 4(a). The total $Q$ factor, defined by $Q = \left(1/Q_0 + 1/Q_{-1}\right)^{-1}$, takes into account the radiative losses from the 0th-order diffraction ($Q_0$) and −1st-order diffraction ($Q_{-1}$). In the upper and lower panels of Fig. 4, $Q_0$ and $Q_{-1}$ are plotted separately as purple and green lines, respectively, and the polarization states of the 0th-and −1st-order diffraction are also shown correspondingly. Since $Q$ diverges at the multi-channel BIC for the thickness $h=1.948a$, both $Q_0$ and $Q_{-1}$ have to diverge simultaneously. First, this implies that there is one topological charge (marked by the black dot) in both two polarization maps as shown in Fig. 4(c); second, the two topological charges coincide with each other in momentum space, giving rise to a multi-channel BIC without any leakage. Note that the topological charge is defined by $v_m = (1/2\pi) \oint_L d\mathbf{k}_\| \cdot \nabla_{\mathbf{k}_\|} \phi_m(\mathbf{k}_\|)$. Here, $L$ is a closed loop in momentum space surrounding the singular point in the counterclockwise direction, and $\phi_m(\mathbf{k}_\|) = 1/2 \arg\left[S_{1,m}(\mathbf{k}_\|) + iS_{2,m}(\mathbf{k}_\|)\right]$ is the orientation angle of the polarization state, where $S_{i,m}$ is the Stokes parameter of the $m$th-order diffraction. For the multi-channel BIC in Fig. 4(c), the topological charges are $v_0 = +1$ and $v_{-1} = -1$. It is worth emphasizing that these two topological charges come from the same eigenstate with fixed $\mathbf{k}_\|$ and $\omega$ but belong to different radiation channels (i.e., the propagating diffraction orders with $\mathbf{k}_\|$ and $\mathbf{k}_\|-\mathbf{G}$). Therefore, they are independent and will not merge or annihilate each other in momentum space. This topological property is distinct from that of merging BIC [13, 31, 38], wherein the topological charges are linked to the same radiation channel.

When the thickness of the PhC slab is slightly varied away from $h_\text{BIC}$, the multi-channel BIC no longer exists, as shown in Figs. 4(b) and 4(d), and $Q_0$ is bounded, whereas $Q_{-1}$ still diverges at a certain $k_x$. The topological charge $v_0 = +1$ for the 0th-order diffraction splits into two half-integer charges of 1/2 with the total topological charge conserved and each being circularly polarized. Because of the $y$-mirror symmetry of the system, the two circularly polarized states are symmetric about the $k_x$ axis and carry the same charge with different handedness (or chirality). The states with right-handed circular polarization (RCP) and left-handed circular polarization (LCP) are indicated by red and blue dots, respectively, in the upper panels of Figs. 4(b) and 4(d). The splitting of an integer charge into two half-integer charges here comes only from the change of thickness and the symmetry of the system is perfectly maintained. Note that below the diffraction limit the breaking of the $C_2$ symmetry is necessary in order to observe this kind of splitting [12]. This non-symmetry-breaking induced splitting manifests the unusual topological nature for multi-channel BICs. For the −1st-order diffraction, the topological charge persists and slightly moves along the $k_x$ axis (see the lower panel in Fig. 4). The dotted arrows in Fig. 4 are a guide for the eyes and indicate the evolution of topological charges. The half-integer charge of RCP (red point) passes through the $k_x$ axis from positive to negative $k_y$, while the one of LCP (blue point) passes through the $k_x$ axis from negative to positive $k_y$. The two half-integer charges meet each other at the $k_x$ axis. Multi-channel BICs lying in the $k_x$–$\omega$ space with only three propagating Bloch modes can be understood as the coincidence of two integer charges in momentum space, one coming from the merging of two half-integer charges and the other being a stable integer charge moving on the $k_x$ axis slowly.

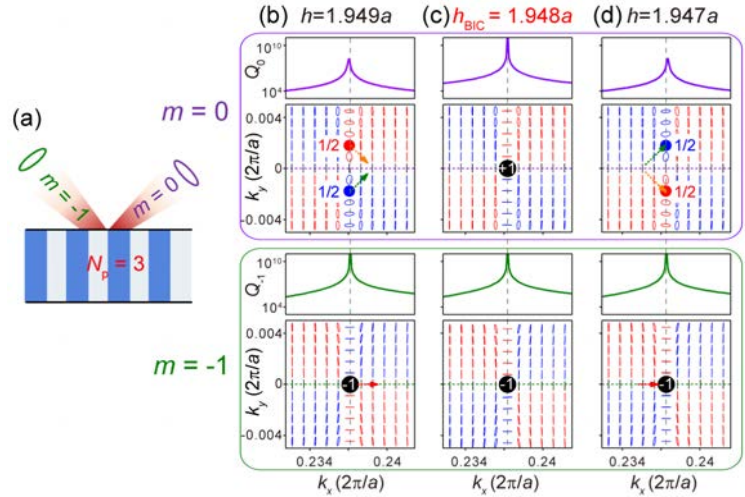

**FIG. 4.** Topological nature of multi-channel BICs. (a) Schematic view of the radiation channels of a guided resonance with $N_r = 2$ (no. of radiation channels) and $N_p = 3$ (no. of propagating Bloch modes). (b)–(d) Evolution of $Q$ factors and polarization maps for different thickness $h$. Results for the 0th- and −1st-order diffraction are shown in the upper and lower panels, respectively. $Q_0$ (purple line) and $Q_{-1}$ (green line) arise from the radiative loss of these two radiation channels. The black (blue and red) dots indexed by the topological charge $\pm 1$ (1/2) represent the vortex centers (circularly polarized states with LCP and RCP). Here, the multi-channel BIC corresponds to that in Fig. 3(a).

Multi-channel BICs can even manifest a different topological nature if they lie in the region of $k_x$–$\omega$ space with different numbers of propagating Bloch modes. Another example, the multi-channel BIC marked in Fig. 3(b), is demonstrated by showing the far-field polarization states of the 0th- and −1st-order diffraction separately in Fig. 5(c). Similarly, both $Q_0$ and $Q_{-1}$ diverge at this BIC point, and the two topological charges coincide with each other in momentum space so that leakage is eliminated for these two radiation channels. Note that the two topological vortexes defined in the two radiation channels can either exhibit the same amount of charge as shown in Fig. 5(c), or different amounts of charge, as shown in Fig. 4(c). Furthermore, both integer charges in Fig. 5(c) will split into a pair of half-integer charges of 1/2 with opposite chirality when the thickness of the PhC slab is slightly varied from $h_{\text{BIC}}$, as shown in Figs. 5(b) and (d). This non-symmetry-breaking induced splitting of an integer charge into two half-integer charges is a generic phenomenon since it happens in both radiation channels. In short, multi-channel BICs in the region with $N_p = 4$ and $N_r = 2$ can also be interpreted as the coincident point of two integer charges in momentum space, both of which result from the merging of two half-integer charges.

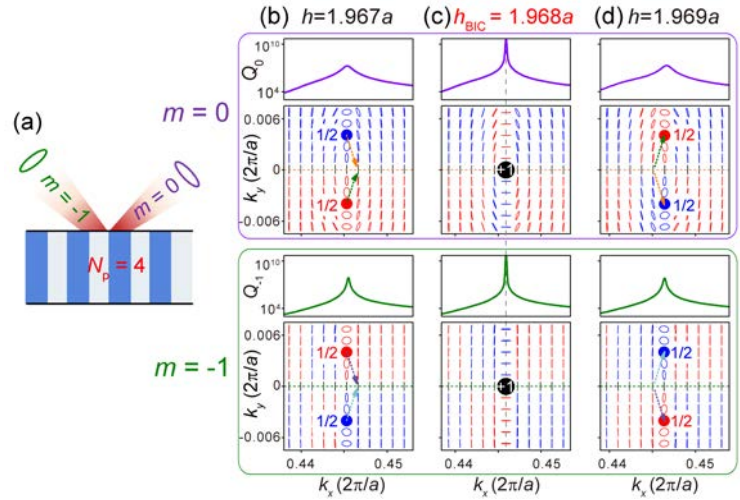

**FIG. 5.** Topological nature of multi-channel BICs. (a) Schematic view of the radiation channels of a guided resonance with $N_r = 2$ (no. of radiation channels) and $N_p = 4$ (no. of propagating Bloch modes). (b)–(d) Evolution of $Q$ factors and polarization maps for different thickness $h$. Results for the 0th- and −1st-order diffraction are shown in the upper and lower panels, respectively. Both integer charges for the 0th and −1st diffraction orders come from the merging of two half-integer charges. The BIC in (c) is the one shown in

Fig. 3(b) in the $k_x$–$\omega$ space.

## CONCLUSION

In summary, we derive the generalized Fresnel equations for the Bloch waves at a PhC/air interface, from which the TIR condition of Bloch waves are obtained. For a PhC slab, by combining the TIR of Bloch waves and the guidance condition, the generalized conditions for waveguide modes are given, with solutions being precisely the BICs. Distinct from BICs below the diffraction limit, multi-channel BICs with frequencies beyond the diffraction limit are found which can only exist for some specific geometric parameters of the PhC slab. They possess a quite different topological nature stemming from the coincidence of two integer charges in the polarization maps of two different radiation channels. Integer topological charges can split into two half-integer charges even without breaking any symmetry, which is generic for multi-channel BICs. Our BIC solver with the generalized conditions for waveguiding in PhC slabs incorporated offers a powerful tool for readily finding BICs at any frequency in momentum space. The revealed distinct topological nature of multi-channel BICs from conventional ones may render new opportunities in designs and applications of BICs possible in nanophotonics and enhanced light-matter interactions as well.

## ACKNOWLEDGMENTS
D. H. thanks Profs. H. Chen, L. Lu, L. Shi and C. Peng for helpful discussions.

## FUNDING
This work was supported by the National Natural Science Foundation of China (grant nos. 12074049, 11727811, 12047564), and Fundamental Research Funds for the Central Universities (grant nos. 2020CDJQY-Z006, 2020CDJQY-Z003).

## AUTHOR CONTRIBUTIONS
D.Z.H. and J.Z. conceived the study and supervised the overall project. P.H., C.W.X., Q.J.S. and D.Z.H. carried out numerical simulations and theoretical analysis. C.W.X., P.H., and A.C. designed the solver. A.C., H.X. and J.Z. contributed to the data analyses and discussions. P.H. and D.Z.H. wrote the paper with revisions from other authors.

*Conflict of interest statement.* None declared.

Peng Hu[1,†], Chongwu Xie[1,†], Qianju Song[1], Ang Chen[2], Hong Xiang[1,3], Dezhuan Han[1,*], and Jian Zi[2,*]

[1] College of Physics, Chongqing University, Chongqing 401331, China

[2] State Key Laboratory of Surface Physics, Key Laboratory of Micro- and Nano-Photonic Structures (Ministry of Education) and Department of Physics, Fudan University, Shanghai 200433, China

[3] Chongqing Key Laboratory for Strongly Coupled Physics, Chongqing 401331, China

**\*Corresponding authors.** Email: dzhan@cqu.edu.cn and jzi@fudan.edu.cn
[†]Equally contributed to this work.

**Dispersion relation for a 1D photonic crystal**

For a binary photonic crystal (PhC), the dispersion relation relates the frequency $\omega$, the normal component of the wave vector $k_z$, and the Bloch wave vector $k_x$. For TE waves, it is given by [1]:

$$\cos k_x a = \cos k_{1x}(a-d)\cos k_{2x}d - \tfrac{1}{2}(\eta+1/\eta)\sin k_{1x}(a-d)\sin k_{2x}d, \quad \text{(S1)}$$

where $\eta = k_{1x}/k_{2x}$, $k_{ix} = \sqrt{\varepsilon_i k_0^2 - k_z^2}$ ($i=1,2$), $k_0 = \omega/c$, and $c$ is the speed of light in free space. For fixed $k_x$ and $\omega$, there are a finite number of propagating states with $k_z^2 > 0$ and an infinite number of evanescent states with $k_z^2 < 0$ in the PhC. The isofrequency contour for the propagating states at a fixed $\omega$ can be obtained by Eq. S1, whereas the frequency ranges with different number of propagating Bloch modes ($N_p$) can be simply obtained by setting $k_z = 0$ in Eq. S1 and counting the number of bands below this frequency. Additionally, the frequency ranges with a different number of radiation channels ($N_r$) in free space can be directly obtained by folding the light line.

**Method used to obtain polarization states of the *m*th-order diffraction**

The electric field **E** and magnetic field **H** of guided resonance in a PhC slab can be simulated by the finite element method. For the guided resonances in the frequency range where there are multiple propagating diffraction orders ($N_r > 1$), multiple propagating plane waves radiate to different directions in free space, and thus the polarization directions of the corresponding far-field radiation cannot be directly defined. It is essential to extract these nonzero propagating-wave amplitudes from the far-field radiation. Using the Bloch theorem, the far-field component of the $m$th-order diffraction can be obtained by $\mathbf{E}_m(\mathbf{k}_\parallel) = 1/a \int_{x_0}^{x_0+a} \mathbf{E} e^{-i(k_x + mG)x} dx$ for the electric field and $\mathbf{H}_m(\mathbf{k}_\parallel) = 1/a \int_{x_0}^{x_0+a} \mathbf{H} e^{-i(k_x + mG)x} dx$ for the magnetic field, where integrations are performed on the horizontal plane away from the PhC slab. Thus, we can evaluate separately the radiation power, $P_m = 2 \int_{x_0}^{x_0+a} \mathbf{S}_m \cdot \hat{z} dx$, per unit cell from the $m$th-order diffraction of the guided resonance, where the multiple of two comes from the symmetry of the structure in the $z$ direction and $\mathbf{S}_m = 1/2 \operatorname{Re}(\mathbf{E}_m \times \mathbf{H}_m^*)$ is the time-averaged Poynting vector. The quality factor that accounts for the radiative loss from the $m$th-order diffraction can then be obtained by $Q_m = \omega U_{\text{eff}} / P_m$, where $U_{\text{eff}}$ is the stored energy in the PhC slab and can be directly calculated by numerical simulations.

To characterize the polarization states of the $m$th-order diffraction, we decompose $\mathbf{E}_m(\mathbf{k}_\parallel)$ into two orthogonal components on the $s$–$p$ plane: $\mathbf{E}_m(\mathbf{k}_\parallel) = t_m^s(\mathbf{k}_\parallel) \hat{e}_m^s + t_m^p(\mathbf{k}_\parallel) \hat{e}_m^p$, where $\hat{e}_m^s = \hat{z} \times \mathbf{k}_m / |\hat{z} \times \mathbf{k}_m|$, $\hat{e}_m^p = \mathbf{k}_m \times \hat{e}_m^s / |\mathbf{k}_m \times \hat{e}_m^s|$, and $\mathbf{k}_m = (k_x + mG)\hat{x} + k_y \hat{y} + k_{z,m} \hat{z}$ is the wave vector of the $m$th-order diffracted wave in free space. Further, the Stokes parameters of $\mathbf{E}_m$ are employed to describe the corresponding polarization states, namely, $S_{0,m} = |t_m^p(\mathbf{k}_\parallel)|^2 + |t_m^s(\mathbf{k}_\parallel)|^2$, $S_{1,m} = |t_m^p(\mathbf{k}_\parallel)|^2 - |t_m^s(\mathbf{k}_\parallel)|^2$, $S_{2,m} = 2 \operatorname{Re}[t_m^{p*}(\mathbf{k}_\parallel) t_m^s(\mathbf{k}_\parallel)]$ and $S_{3,m} = 2 \operatorname{Im}[t_m^{p*}(\mathbf{k}_\parallel) t_m^s(\mathbf{k}_\parallel)]$.

**Some examples of the BIC solver**

Based on the generalized conditions for waveguide modes that are explained in the main text, a BIC solver for 1D PhC slab is developed. Here, we first show an example of the BIC solver in the frequency range with $N_p = 2$ and $N_r = 1$, which is indicated by the gray shaded region in **Supplementary Fig. 1a**. Assuming that the total internal reflection (TIR) of TE Bloch waves is satisfied at the upper interface of the PhC slab and $N$ modes are considered in total, by eliminating of the coefficients $t_m$, Eq. 6 in the main text can be rewritten as follows:

$$\sum_{n=1}^{N}(a_n e^{ik_z^{(n)}h/2}\alpha_{m,n} + r_n e^{-ik_z^{(n)}h/2}\beta_{m,n}) = 0, \tag{S2}$$

where $\alpha_{m,n} = (1-\rho_{m,n})\tilde{u}_m^{(n)}$, $\beta_{m,n} = (1+\rho_{m,n})\tilde{u}_m^{(n)}$, $\rho_{m,n} = k_z^{(n)}/k_{z,m}$ and $\tilde{u}_m^{(n)} = \vec{X}_{mn}$. The TIR condition in Eq. 9 in the main text can be expressed as follows:

$$\sum_{n=1}^{N}(a_n e^{ik_z^{(n)}h/2} \pm r_n e^{-ik_z^{(n)}h/2})\tilde{u}_0^{(n)} = 0. \tag{S3}$$

Here, the origin of the $z$ axis is set at the center of the PhC slab for convenience and one more equation in Eq. S2 is required if one more evanescent mode is taken into account.

Given an initial $h$, the reflection phase shift $\varphi_r^{(n)} = \arg(r_n/a_n e^{-ik_z^{(n)}h})$ at the upper interface for every ($k_x$, $\omega$) point can be obtained by solving Eqs. S2 and S3 with $r_n = a_n$ ($n > 3$) for all evanescent modes. In the case of $N_p - N_r = 1$, the phase shifts $\varphi_r^{(n)}$ are fixed at any ($k_x$, $\omega$) point for different thickness $h$. The phase shift of the first Bloch wave $\varphi_r^{(1)}$ is taken as example and shown in **Supplementary Fig. 1b** for a fixed $k_x$ and a different number of evanescent Bloch waves. It is found that the convergence is very fast as the number of evanescent Bloch waves $N - N_p \geq 2$. In other words, the positions of BICs in the $k_x$–$\omega$ space converge very quickly if only a few evanescent waves are considered in addition to the propagating Bloch waves. For simplicity, we choose $N - N_p = 2$ in the solver by default to find BICs. In this way, a database of $\varphi_r^{(n)}$ for a PhC in the whole $k_x$–$\omega$ space can be built. Finally, for any thickness $h$, the total phase of a round trip for the $n$th propagating Bloch wave inside the PhC slab is simply $k_z^{(n)}h + \varphi_r^{(n)}$. What the solver should do is to determine whether this phase is integer multiples of $\pi$. Moreover, when the electric field **E** and magnetic field **H** as well as $\varepsilon$ and $\mu$ are simultaneously exchanged, Maxwell's equations will remain unchanged [2], and thus the solver can also be directly applied to TM Bloch waves. When other structural parameters of a PhC slab are fixed, the obtained TE and TM BICs for different thickness $h$ (solid lines) are shown in **Supplementary Fig. 1c and 1d**, respectively, agreeing well with the simulated results

(dots). We further show the calculation process for the TE case in **Supplementary Video 1**, fully demonstrating the low computational complexity of this solver.

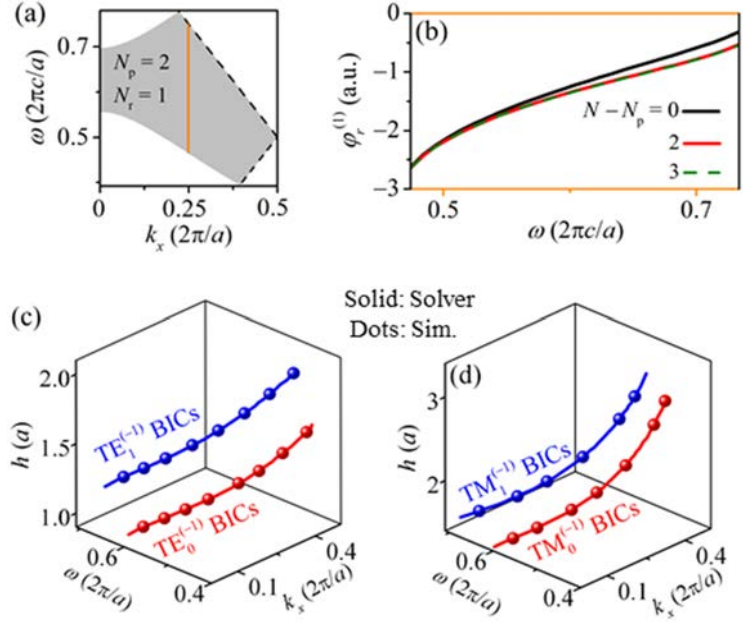

**Supplementary Fig. 1** Examples of the BIC solver in the frequency range with $N_p = 2$ and $N_r = 1$. (a) The gray shaded region indicates the frequency range, and the black dashed lines represent the folded light line in free space. (b) Convergence of the phase shift of the first Bloch wave $\varphi_r^{(1)}$ for different numbers of evanescent Bloch waves $N - N_p$. This example is calculated at $k_x = 0.5\pi/a$, indicated by the orange line in (a). (c) and (d) TE and TM BICs for different thickness $h$. Solid lines denote the results of the BIC solver, whereas the simulated results are denoted by dots. Here, the other structural parameters are chosen as $\varepsilon_1 = 1$, $\varepsilon_2 = 4.9$, and $d = 0.5a$.

The generalized conditions for waveguide modes in PhC slabs can be applied to not only the $k_x$ axis but also the whole Brillouin zone. Therefore, this BIC solver can work in the whole $\mathbf{k}_\parallel$–$\omega$ space, where $\mathbf{k}_\parallel = (k_x, k_y)$. For a 1D PhC, when $k_y$ is not equal to zero, the bulk mode is no longer a pure TE or TM Bloch wave. Thus, for the guided resonances on the TE band in the $k_x$ axis consisting of two TE propagating Bloch waves, the guided resonances on the corresponding TE-like band are composed of four propagating Bloch waves, i.e., two TE and two TM propagating Bloch waves, and $N_p = 4$. Moreover, these guided resonance for $k_y \neq 0$ radiate to far field by both the $s$- and $p$-polarization channels, namely, $N_r = 2$. In this case ($N_p - N_r = 2$), to find a convergent solution, an additional condition, $r_n = a_n$ for one of the four propagating Bloch modes, is also adopted in addition to $r_n = a_n$ for all evanescent Bloch modes. In this solver, we choose $r_1 = a_1$ for the first propagating Bloch mode by default. However, this additional condition makes the phase shift $\varphi_r^{(n)}$ at a certain ($\mathbf{k}_\parallel$, $\omega$) point no longer a constant for different $h$ when the TIR condition of Bloch waves is satisfied at the upper interface. Therefore, the solver has to recalculate the phase shift $\varphi_r^{(n)}$ for different $h$ to find BICs. Generally, BICs exist on the high symmetry lines, and an example of determination of BICs in the $k_y$ axis in the parameter space is shown in **Supplementary Fig. 2**, which is also in full agreement with the simulated results.

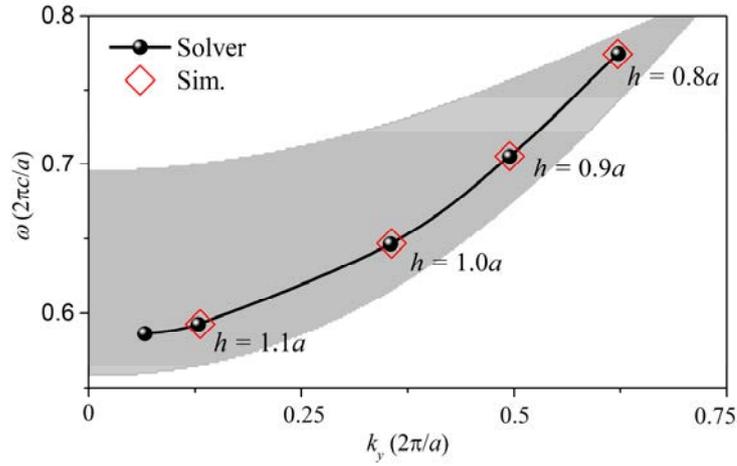

**Supplementary Fig. 2** Examples of the BIC solver in the $k_y$ axis. The gray shaded region indicates the region with $N_p = 4$ and $N_r = 2$. Solid line and black dots denote the results of the BIC solver, and red rhombuses indicated the simulated results for different $h$.

Furthermore, the generalized conditions for waveguide modes can be directly applied to the case with multiple radiation channels and thus the BIC solver can also work well to search multi-channel BICs. For the case of two radiation channels, the BIC solver can readily find multi-channel BICs in a wide range of thickness $h$ with fast convergence, as shown in **Supplementary Fig. 3**. Similar to that shown in **Supplementary Fig. 1b**, the results converge very quickly when a few evanescent waves are considered. Examples of multi-channel BICs with $h_{BIC}$ less than $3a$ are shown in **Supplementary Fig. 3b**, and the corresponding parameters ($h_{BIC}$, $k_{x,BIC}$, $\omega_{BIC}$) are also exhibited in **Supplementary Table 1** for clarity. The multi-channel BICs labelled by ① and ⑤ correspond to those in Fig. 3(a) and 3(b) in the main text, respectively.

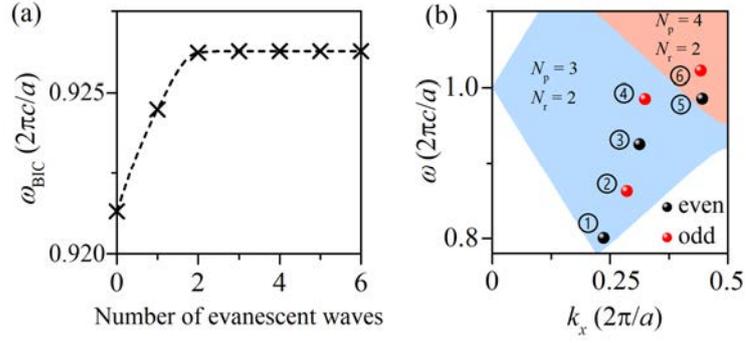

**Supplementary Fig. 3.** BICs with two radiation channels. (a) Convergence of $\omega_{BIC}$ for different number of evanescent modes. (b) Examples of multi-channel BICs with $h_{BIC} < 3a$. The other system parameters are chosen as $\varepsilon_1 = 1$, $\varepsilon_2 = 4.9$, and $d = 0.5a$.

**Supplementary Table 1.** Examples of multi-channel BICs for $h_{BIC} < 3a$.

| No. | $h_{BIC}$ ($a$) | $k_{x,BIC}$ ($2\pi/a$) | $\omega_{BIC}$ ($2\pi c/a$) |
| --- | --- | --- | --- |
| ① | 1.948 | 0.237 | 0.800 |
| ② | 2.069 | 0.286 | 0.862 |
| ③ | 2.168 | 0.312 | 0.926 |
| ④ | 2.261 | 0.325 | 0.985 |
| ⑤ | 1.968 | 0.446 | 0.985 |
| ⑥ | 2.147 | 0.443 | 1.022 |

When there exist more radiation channels in free space ($N_r \geqslant 3$), the construction of a BIC requires that all Q factors accounting for the radiative loss towards different radiation channels, i.e., $Q_0$, $Q_{\pm 1}$, …, should diverge at the same point in momentum space simultaneously. This is very difficult to achieve since one needs many degrees of freedom to cancel out all these losses in different channels and search in a very large parameter space. In this sense, an algorithm with very fast convergence speed is especially useful. As examples, by only varying the thickness $h$, the $Q$ factors of quasi-BICs embedded in multiple radiation channels can be maximized very quickly. A quasi-BIC for $h = 13.018a$ with 3 radiation channels is shown in **Supplementary Fig. 4.** This multi-channel quasi-BIC appear on the band of the PhC slab, as highlighted by red dots in **Supplementary Fig. 4a**. The corresponding $Q$ factors of this band are plotted in **Supplementary Fig. 4b**, where the inset shows a zoomed-in figure near the quasi-BIC. Similar to Figs. 4 and 5 in the main text, the corresponding polarization maps (left panel) and $Q$ factors (right panel) for the 0th- (upper), −1st- (middle), and 1st- (lower) order diffractions in the vicinity of quasi-BIC are shown in **Supplementary Fig. 4c**. Clearly, only $Q_1$ diverges at a certain $k_x$, corresponding to one topological charge (black dot) in the polarization map for the 1st-order diffraction. The peaks of $Q_0$, $Q_{-1}$ and $Q_1$ still deviate from each other slightly in the $k_x$ axis since here we only optimize the slab thickness. More degrees of freedom are needed to be considered to achieve an ideal BIC. For comparison, the $Q$ factor for the same band with $h = 13.1a$ is also shown in **Supplementary Fig. 4b**, and there are no quasi-BICs with ultra-high $Q$ factor.

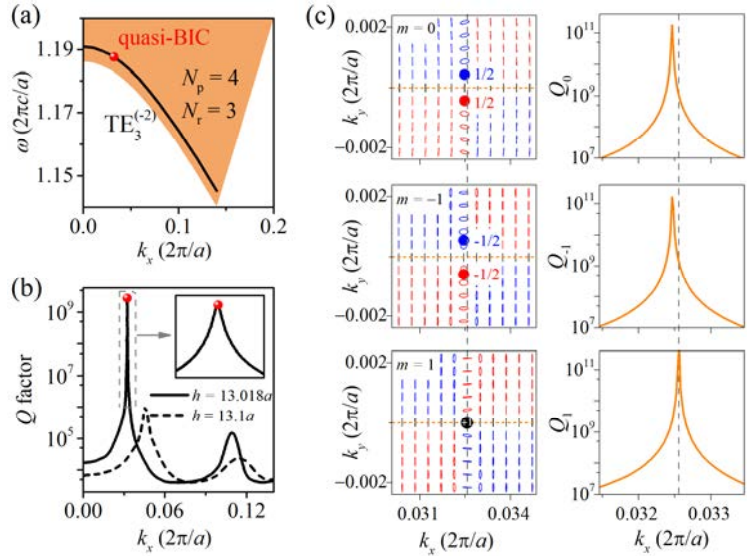

**Supplementary Fig. 4.** Quasi-BICs with three radiation channels. (a) Simulated band structure with $h = 13.018a$. A multi-channel quasi-BIC (red dot) exists on the band. The orange shaded region indicates the region in which there are four propagating Bloch modes ($N_p = 4$) in the photonic crystal and three radiation channels ($N_r = 3$) in free space. (b) Simulated $Q$ factors of the guided resonances for $h = 13.018a$ (solid line) and $h = 13.1a$ (dashed line) in the $TE_3^{(-2)}$ band. (c) Polarization maps (left panel) and $Q$ factors (right panel) for 0th- (upper), −1st- (middle) and 1st- (lower) order diffractions in the vicinity of the quasi-BIC at $h = 13.018a$. Black (blue and red) dots indexed by the topological charge $\pm 1$ ($\pm 1/2$) represent the vortex centers (circularly polarized states with LCP or RCP). The half charges lie very close to but still away from the $k_x$ axis when the $Q$ factor is maximized by varying $h$. The other system parameters are chosen as $\varepsilon_1 = 1$, $\varepsilon_2 = 4.9$, and $d = 0.5a$.

### Total internal reflection of two Bloch waves

For the PhC with a small periodic index modulation, we assume that only propagating Bloch waves are considered and all evanescent Bloch waves are excluded here. We consider the case when there are two propagating Bloch waves ($N_p = 2$) and only one radiation channel in free space ($N_r = 1$). For the $n$th TE Bloch wave in the PhC, the periodic-in-cell part of the electric field can be expanded as $u^{(n)}(x) = 1 + \tilde{u}_{-1}^{(n)} e^{-iGx} + \cdots$, where the 0th Fourier coefficient is specially set to be one. When the two TE Bloch waves impinge on the PhC interface, for $m = 0$, the boundary conditions in Eqs. 4 and 5 in the main text can be rewritten as follows:

$$\sum_{n=1}^{2} (a_n + r_n) = t_0 \tag{S4}$$

and

$$\sum_{n=1}^{2} (a_n - r_n) k_z^{(n)} = k_{z,0} t_0. \tag{S5}$$

For $m = -1$, by eliminating coefficients $t_m$ of the diffracted evanescent waves in free space, we have the following:

$$\sum_{n=1}^{2} \left( a_n \left( k_{z,-1} - k_z^{(n)} \right) + r_n \left( k_{z,-1} + k_z^{(n)} \right) \right) \tilde{u}_{-1}^{(n)} = 0. \tag{S6}$$

When the total internal reflection (TIR) of two TE Bloch waves occurs, namely, $t_0 = 0$ and by combining Eqs. S4–S6, the relative incidence coefficient can be obtained and written as follows:

$$\frac{a_2}{a_1} = -\frac{1 + Z_2}{1 + Z_1}, \tag{S7}$$

where $Z_n = k_{z,-1} / k_z^{(n)}$ for $n$th propagating TE Bloch waves. The expression for the reflection coefficient is written as follows:

$$\frac{r_n}{a_n} = \frac{1 - Z_n}{1 + Z_n}. \tag{S8}$$

These are Eqs. 11 and 12 in the main text.

For the TM Bloch waves, the eigen wave function changes from electric to magnetic field, which can be expressed as $H_y^{(n)}(x) = v^{(n)}(x) e^{i(k_x x + k_z^{(n)} z)}$ for the $n$th state. Similar to the TE Bloch waves, the periodic part can be expanded as $v^{(n)}(x) = 1 + \tilde{v}_{-1}^{(n)} e^{-iGx} + \tilde{v}_1^{(n)} e^{iGx} + \cdots$. In this case, the periodic index modulation $\varepsilon(x)$ plays an important role in the boundary condition, and we adopt the first-order approximation for the Fourier transform of $\varepsilon^{-1}(x)$, that is, $\varepsilon^{-1}(x) \approx \kappa_0 + \kappa_1 e^{iGx} + \kappa_{-1} e^{-iGx}$. The wave equation for $H_y^{(n)}(x)$ is given by:

$$\frac{\partial \varepsilon^{-1}(x)}{\partial x} \frac{\partial H_y^{(n)}}{\partial x} + \varepsilon^{-1}(x) \left( \frac{\partial^2}{\partial x^2} + \frac{\partial^2}{\partial z^2} \right) H_y^{(n)} = -k_0^2 H_y^{(n)}. \tag{S9}$$

Substituting the expansion of $H_y^{(n)}$ and $\varepsilon^{-1}(x)$ into this wave equation, we obtain

$$\kappa_1 \left\{ (k_x + mG)(k_x + (m-1)G) + \left(k_z^{(n)}\right)^2 \right\} \tilde{v}_{m-1}^{(n)} + \kappa_{-1} \left\{ (k_x + mG)(k_x + (m+1)G) + \left(k_z^{(n)}\right)^2 \right\} \tilde{v}_{m+1}^{(n)}$$
$$\approx \left\{ k_0^2 - \kappa_0 \left[ (k_x + mG)^2 + \left(k_z^{(n)}\right)^2 \right] \right\} \tilde{v}_m^{(n)}. \tag{S10}$$

For $m=0$,

$$\tilde{v}_0^{(n)} \approx \frac{\kappa_1 \left[ k_x(k_x - G) + \left(k_z^{(n)}\right)^2 \right] \tilde{v}_{-1}^{(n)} + \kappa_{-1} \left[ k_x(k_x + G) + \left(k_z^{(n)}\right)^2 \right] \tilde{v}_1^{(n)}}{k_0^2 - \kappa_0 \left[ k_x^2 + \left(k_z^{(n)}\right)^2 \right]}. \tag{S11}$$

For $m=-1$,

$$\tilde{v}_{-1}^{(n)} \approx \frac{\kappa_1 \left[ (k_x - G)(k_x - 2G) + \left(k_z^{(n)}\right)^2 \right] \tilde{v}_{-2}^{(n)} + \kappa_{-1} \left[ k_x(k_x - G) + \left(k_z^{(n)}\right)^2 \right] \tilde{v}_0^{(n)}}{k_0^2 - \kappa_0 \left[ (k_x - G)^2 + \left(k_z^{(n)}\right)^2 \right]}. \tag{S12}$$

For a vanishing index modulation ($\Delta = (\varepsilon_2 - \varepsilon_1)/\varepsilon_1 \rightarrow 0$), $k_z^{(1)}$ and $k_z^{(2)}$ approach $\sqrt{\varepsilon_{\text{eff}} k_0^2 - (k_x - G)^2}$ and $\sqrt{\varepsilon_{\text{eff}} k_0^2 - k_x^2}$, respectively, where $\varepsilon_{\text{eff}} \approx 1/\kappa_0$ is the effective permittivity for the TM case. Therefore, if $k_x$ is not near the Brillouin zone center, $\tilde{v}_{-1}^{(1)}$ is dominant in the expansion of $H_y^{(1)}$, whereas the dominant term of $H_y^{(2)}$ is $\tilde{v}_0^{(2)}$. In this case, we only keep $\tilde{v}_0^{(n)}$ and $\tilde{v}_{-1}^{(n)}$ in the expansion of $H_y^{(n)}$ and neglect all other terms. We can then obtain $\tilde{v}_{-1}^{(n)}$ from Eqs. S11 and S12, that is:

$$\tilde{v}_{-1}^{(1)} \approx \frac{k_0^2 - \kappa_0 \left[ k_x^2 + \left(k_z^{(1)}\right)^2 \right]}{\kappa_1 \left[ k_x(k_x - G) + \left(k_z^{(1)}\right)^2 \right]} \quad \text{and} \quad \tilde{v}_{-1}^{(2)} = \frac{k_0^2 - \kappa_0 \left[ k_x^2 + \left(k_z^{(2)}\right)^2 \right]}{\kappa_1 \left[ k_x(k_x - G) + \left(k_z^{(2)}\right)^2 \right]} \approx 0. \tag{S13}$$

When the TIR of two TM Bloch waves occurs at the PhC interface, the boundary conditions for $m=0$ are as follows:

$$\sum_{n=1}^{2} (a_n + r_n) = 0 \tag{S14}$$

and

$$\sum_{n=1}^{2} k_z^{(n)} \left( a_n - r_n \right) \left( \kappa_0 + \kappa_1 \tilde{v}_{-1}^{(n)} \right) = 0 . \tag{S15}$$

Similar to Eq. S6, for $m=-1$, we have the following:

$$\sum_{n=1}^{2} a_n \left( k_{z,-1} \tilde{v}_{-1}^{(n)} / \varepsilon_b - k_z^{(n)} \left( \kappa_{-1} + \kappa_0 \tilde{v}_{-1}^{(n)} \right) \right) + r_n \left( k_{z,-1} \tilde{v}_{-1}^{(n)} / \varepsilon_b + k_z^{(n)} \left( \kappa_{-1} + \kappa_0 \tilde{v}_{-1}^{(n)} \right) \right) = 0 . \tag{S16}$$

Combining Eqs. S14–S16, the relative incidence coefficient and reflection coefficients for the TIR of two TM Bloch waves are derived and, respectively, expressed as follows:

$$\frac{a_2}{a_1} = -\frac{k_z^{(1)} \left[ \varepsilon_b k_z^{(2)} \left( \kappa_0^2 - \kappa_{-1} \kappa_1 \right) + k_{z,-1} \left( \kappa_1 \tilde{v}_{-1}^{(1)} + \kappa_0 \right) \right]}{k_z^{(2)} \left[ \varepsilon_b k_z^{(1)} \left( \kappa_0^2 - \kappa_{-1} \kappa_1 \right) + k_{z,-1} \left( \kappa_1 \tilde{v}_{-1}^{(2)} + \kappa_0 \right) \right]}, \tag{S17}$$

$$\frac{r_1}{a_1} = \frac{\varepsilon_b k_z^{(1)} \left( \kappa_0^2 - \kappa_{-1} \kappa_1 \right) - k_{z,-1} \left( \kappa_1 \tilde{v}_{-1}^{(2)} + \kappa_0 \right)}{\varepsilon_b k_z^{(1)} \left( \kappa_0^2 - \kappa_{-1} \kappa_1 \right) + k_{z,-1} \left( \kappa_1 \tilde{v}_{-1}^{(2)} + \kappa_0 \right)}, \tag{S18}$$

and

$$\frac{r_2}{a_2} = \frac{\varepsilon_b k_z^{(2)} \left( \kappa_0^2 - \kappa_{-1} \kappa_1 \right) - k_{z,-1} \left( \kappa_1 \tilde{v}_{-1}^{(1)} + \kappa_0 \right)}{\varepsilon_b k_z^{(2)} \left( \kappa_0^2 - \kappa_{-1} \kappa_1 \right) + k_{z,-1} \left( \kappa_1 \tilde{v}_{-1}^{(1)} + \kappa_0 \right)} . \tag{S19}$$

Substituting Eq. S13 into Eqs. S17–S19 and neglecting the term $\kappa_{-1} \kappa_1$, they can be simplified as $\frac{a_2}{a_1} = -\frac{1+Z_2}{1+Z_1}$ and $\frac{r_n}{a_n} = \frac{1-Z_n}{1+Z_n}$, where $Z_1 = \frac{k_{z,-1}/\varepsilon_b}{k_z^{(1)}/\varepsilon_H}$ and $Z_2 = C \frac{k_{z,-1}/\varepsilon_b}{k_z^{(2)}/\varepsilon_H}$ with $\varepsilon_H = \kappa_0^{-1}$ and $C = \left( \left( k_z^{(2)} \right)^2 + k_x^2 - k_x G \right) / \left( \left( k_z^{(2)} \right)^2 + k_x^2 + k_x G - G^2 \right)$, corresponding to Eq. 13 in the main text. Here, we use the relation $\left( k_z^{(2)} \right)^2 \approx \left( k_z^{(1)} \right)^2 - G^2 + 2 k_x G$ for small index contrast in the expression of $C$.